\documentclass[showpacs,preprintnumbers,amsmath,amssymb]{revtex4}

\usepackage{graphicx}
\usepackage{dcolumn}
\usepackage{bm}
\usepackage{mathrsfs}
\usepackage{amssymb}
\usepackage{amsmath}
\usepackage{amsfonts}
\usepackage{amsfonts,enumerate}
\usepackage{pifont}
\usepackage{enumerate}
\usepackage{graphics}
\usepackage{verbatim}
\usepackage{amsthm}
\numberwithin{equation}{section}

\begin{document}


\title{Mathematical formalism of many-worlds quantum mechanics}

\author{Zeqian Chen}
\affiliation{%
State Key Laboratory of Resonances and Atomic and Molecular Physics, Wuhan Institute of Physics and Mathematics, Chinese Academy of Sciences, 30 West District, Xiao-Hong-Shan, Wuhan 430071, China}%


\begin{abstract}
We combine the ideas of Dirac's orthonormal representation, Everett's relative state, and 't Hooft's ontological basis to define the notion of a world for quantum mechanics. Mathematically, for a quantum system $\mathcal{Q}$ with an associated Hilbert space $\mathbb{H},$ a world of $\mathcal{Q}$ is defined to be an orthonormal basis of $\mathbb{H}.$ The evolution of the system is governed by Schr\"{o}dinger's equation for the worlds of it. An observable in a certain world is a self-adjoint operator diagonal under the corresponding basis. Moreover, a state is defined in an associated world but can be uniquely extended to the whole system as proved recently by Marcus, Spielman, and Srivastava. Although the states described by unit vectors in $\mathbb{H}$ may be determined in different worlds, there are the so-called topology-compact states which must be determined by the totality of a world. We can apply the Copenhagen interpretation to a world for regarding a quantum state as an external observation, and obtain the Born rule of random outcomes. Therefore, we present a mathematical formalism of quantum mechanics based on the notion of a world instead of a quantum state.
\end{abstract}

\maketitle

\section{Introduction}\label{Intro}

In the conventional formulation of quantum mechanics as in \cite{vN1955}, a system is associated with a Hilbert space $\mathbb{H},$ and this system is completely described by a (pure) state $\psi,$ which is a normalized element of $\mathbb{H},$ in the sense that it gives maximal information that external observers can obtain by specifying the probabilities of the results of various observation which can be made on the system. Thus the meaning of a quantum state is explained by ``external observation" in terms of classical instrument. This is the key content of the Copenhagen interpretation for the standard quantum mechanics based on the notion of a quantum state. Objections to this ``external observation" formulation of quantum mechanics have been raised by some founders of quantum theory, among them there are Einstein and Schr\"{o}dinger, well known as the EPR paradox \cite{EPR} and Schr\"{o}dinger's cat \cite{SchrodingerCat} respectively. Mathematically speaking, the ``external observation" formulation is a statistical description of a quantum system based on observation in terms of classical instrument.

In 1950s', Everett \cite{Everett1957} presented a ``relative state" formulation of quantum mechanics, now well known as the many-worlds interpretation (MWI) of quantum mechanics (cf. \cite{Vaidman2002}). In this formulation, a wave function that obeys the Schr\"{o}dinger equation everywhere and at all times supplies a complete mathematical model for every isolated physical system without exception. The main point of the MWI different from the conventional interpretation is that the external observation is made by observer obeying quantum mechanics that can be regarded as part of a larger isolated system. As a consequence, in Everett's theory, the measurement problem reduces to that all outcomes of measurement should appear at the same instant but in different ``worlds", and so no wavepacket collapse occurs, contrary to the orthodox formulation \cite{Dirac1958}. However, the concept of a world in the MWI is not a rigorously defined entity, but expressed as a term in the orthonormal decomposition of the quantum state of measurement instrument. Yet, the MWI is primitively based on the notion of a quantum state as the same as the conventional formulation of quantum mechanics, whereas the world is a derived concept based on the notion of the quantum state \cite{Vaidman2014}.

In this paper, we will develop a realistic description of a quantum system based on the notion of a world. Different from Everett's theory, we give a rigorously mathematical definition of a world, and our formalism is primitively based on the notion of a world instead of a quantum state, whereas the concept of a state is a derived notion as defined in a certain world. Precisely, we define a world to be an orthonormal basis. This mathematical definition of a world for quantum mechanics is motivated by Dirac's orthonormal representation, 't Hooft's ontological basis \cite{Hooft2014}, and the more recent resolution of the Kadison-Singer problem \cite{KS1959} on the unique extension of pure states by Marcus, Spielman, and Srivastava \cite{MSS2015}.

In fact, Dirac has involved an orthonormal basis to characterize a compatible family of physical observables, that is, the family of self-joint operators diagonal under the basis. Such a basis is called an orthonormal representation for the corresponding compatible family of physical observables. This motivated the study of Kadison and Singer about the problem of the unique extension of pure states in 1950s', which was completely solved by Marcus, Spielman, and Srivastava in 2013. On the other hand, t' Hooft recently utilized the notion of an ontological basis to handle his theory of deterministic quantum mechanics, which is postulated as a very special basis for a quantum system. Combing together those works of Dirac, t' Hooft, and Marcus, Spielman and Srivastava shows that the notion of an orthonormal basis should be an essential block for quantum mechanics and could be, in place of the concept of a quantum state, used to build a mathematical foundation of quantum mechanics. This stimulates us to use the basic assumption of an orthonormal basis describing a world for expressing Everett's idea of many worlds. We remark that, in the sense of \cite{HS2010}, Everett's notion of a world in the MWI is an operational concept, while ours is realistic.

In Section \ref{manyworlds}, we present the details of mathematical definitions of a world, observable, and state. In our formalism, the concept of observable is a relative notion, that is to say, one can only ask the observable relative to a certain world of the system. However, the notion of a state has absolute meaning in this formalism, although it must be mathematically defined in a certain world. In particular, the so-called topology-compact states appear naturally in a world except for vector states in the usual formulation of quantum mechanics. As for measurement, we apply the Copenhagen interpretation to a world for regarding a quantum state as an external observation, and obtain the Born rule. That is to say, a quantum state represents the information possessed by an external observer which satisfies the Markovian property. This excludes the occurrence of the wavepacket collapse. In Section \ref{nonlocality}, we use the formalism of quantum mechanics in terms of worlds to discuss how to remove the action at a distance appearing in the conventional formulation of the EPR paradox and various Bell's theorems, in a similar way as in \cite{Tipler2014, Vaidman2015}. In Section \ref{topologystate}, we give a mathematical method of constructing topology-compact states by using the so-called Banach limit. Finally, in Section \ref{Conclusion}, we summarize the results obtained in the previous sections and make some comments on the relationship between our formulation and the conventional one of Everett's many-worlds theory.

\section{Mathematical foundation of quantum mechanics based on the notion of a world}\label{manyworlds}

Our mathematical formulation of many-worlds quantum mechanics is primitively based on the notion of a world, as such the concepts of observable and state are derived notions. For clarity, we divide this section into several subsections.

\subsection{Notion of a world}

Let us consider a quantum system $\mathcal{Q}$ with an associated Hilbert space $\mathbb{H}.$ In what follows, we always assume that $\mathbb{H}$ is a separable Hilbert space with dimension great than one. Mathematically, {\it a world} of $\mathcal{Q}$ is defined to be an orthonormal basis of $\mathbb{H},$ that is a complete orthonormal set of vectors $(e_n)$ in $\mathbb{H}$ such that
\begin{enumerate}[{\rm 1)}]

\item $e_n$'s are all vectors of norm one,

\item $\langle e_j | e_k \rangle =0$ whenever $j \not= k,$ and

\item for any $u \in \mathbb{H},$ we have $u = \sum_n \langle e_n | u \rangle e_n$ in $\mathbb{H}.$
\end{enumerate}
There are many different worlds for $\mathcal{Q},$ thanks to many orthonormal bases existing for a Hilbert space $\mathbb{H}.$ We identify a world $W = (e_n)$ with another one $W' = (e'_n)$ whenever $e_n = \alpha_n e'_n$ for all $n,$ where every $\alpha_n$ is a complex number such that $|\alpha_n| =1.$

It is known that any operator transforming an orthonormal basis into another one is a unitary operator. Conversely, a unitary operator preserves any orthonormal basis, that is, for a given unitary operator $U,$ if $W = (e_n)$ is an orthonormal basis then $U W = (U e_n)$ is again an orthonormal basis. Then, the evolution of worlds satisfies that $U_{t + s} W = U_t U_s W$ for all $t,s \ge 0,$ where $W$ is the initial world, i.e., $W_0 =W,$ and $U_t$'s are all unitary operators. Thus,
\begin{equation}\label{eq:EquaWorld}
\frac{d W_t}{d t} = \mathrm{i} A W_t, \quad t >0
\end{equation}
with $W_0 = W,$ where $A$ is a self-joint operator such that $W_t = U_t W$ with $U_t = e^{\mathrm{i} t A}$ for all $t \ge 0.$ This is the Schr\"{o}dinger equation for the evolution of worlds.

For a composite system $\mathcal{Q}$ composed of two subsystems $\mathcal{Q}_A$ and $\mathcal{Q}_B$ with associated Hilbert spaces $\mathbb{H}_A$ and $\mathbb{H}_B,$ denoted by $\mathcal{Q} = \mathcal{Q}_A \times \mathcal{Q}_B,$ the Hilbert space for $\mathcal{Q}$ is taken to be $\mathbb{H} = \mathbb{H}_A \otimes \mathbb{H}_B,$ the tensor product of $\mathbb{H}_A$ and $\mathbb{H}_B.$ We define a world of $\mathcal{Q} = \mathcal{Q}_A \times \mathcal{Q}_B$ to be an orthonormal basis of $\mathbb{H}$ of product form $(e_n) = (e^A_n \otimes e^B_n),$ where $e^A_n$ and $e^B_n$ are all normalized elements of $\mathbb{H}_A$ and $\mathbb{H}_B$ respectively. Thus for a world $W$ of $\mathcal{Q}_A \times \mathcal{Q}_B,$ there exist a world $W_A = (e^A_n)$ of $\mathcal{Q}_A$ and a world $W_B = (e^B_k)$ of $\mathcal{Q}_B$ such that $W = (e^A_n \otimes e^B_k).$ Note that there are some orthonormal bases of $\mathbb{H} = \mathbb{H}_A \otimes \mathbb{H}_B$ which cannot be of product form and hence does not define a world for $\mathcal{Q} = \mathcal{Q}_A \times \mathcal{Q}_B$ when considered as a composite system, but still defines a world of $\mathcal{Q}$ as a single system.

In general, we can define the notion of a world for multipartite systems in a similar way.

\subsection{Observable}

In a certain world $W = (e_n),$ {\it an observable} is defined to be a self-adjoint operator on $\mathbb{H}$ diagonal under the basis $(e_n).$ The set of all such operators is denoted by $\mathcal{O} (W).$ Note that we are thus faced with a fundamental relativity of observable, which is based on the formalism of quantum mechanics in terms of worlds. It is meaningless to ask an observable for a system under consideration as in the conventional formulation of quantum mechanics, but one needs to ask the observable relative to a certain world of the system.

The relativity of an observable is implicit in Dirac's idea of orthonormal representation \cite[\S 14]{Dirac1958}. In fact, he defined an orthonormal representation to be an orthonormal basis, and further defines a complete set of commuting observables to be a set of observables within which all commute with one another and for which there is only one simultaneous eigenstate belonging to any set of eigenvalues. Furthermore, Dirac pointed out that an orthonormal representation determines a complete set of commuting observables, and vice versa. Since any observable can be made into a complete commuting set by adding certain observales to it, this implies that each observable can be considered in an orthonormal representation. Note that whenever an orthonormal representation is fixed, all the observables relative to it are determined. Hence, we may consider any observable as a relative entity with respect to an orthonormal representation or a world as named above.

We emphasize that the relativity of observables was explicitly raised by 't Hooft in his theory of deterministic quantum mechanics \cite[\S 9]{Hooft2014}. In deterministic quantum mechanics, the existence of a very special basis called the ontological basis is postulated, which is an orthonormal basis and usually related with a symmetry transformation of the system under consideration. t' Hooft pointed out that what really matters is that an ontological basis allows a meaning subset of observables to be defined as operators that are diagonal in this basis. This means that we only need to ask observables that are diagonal in the ontological basis. In our formulation, we generalize the relativity of observables with respect to the ontological basis to any orthonormal basis, and obtain the relativity of any observable with discrete eigenvalues.

Given a world $W = (e^A_n \otimes e^B_k)$ of a composite system $\mathcal{Q} = \mathcal{Q}_A \times \mathcal{Q}_B$ with the associated Hilbert space $\mathbb{H} = \mathbb{H}_A \otimes \mathbb{H}_B,$ an observable in $W$ is by definition a self-adjoint operator $O$ on $\mathbb{H}$ diagonal under the basis $(e^A_n \otimes e^B_k).$ Then we can write such an observable as $O = O_A \otimes O_B,$ where $O_A$ and $O_B$ are respectively observables in $W_A = (e^A_n)$ and $W_B = (e^B_k)$ (notice that such $O_A$ and $O_B$ may be not unique). Conversely, if $O_A$ is an observable in a world $W_A = (e^A_n)$ of $\mathcal{Q}_A$ and $O_B$ in a world $W_B = (e^B_k)$ of $\mathcal{Q}_B,$ then $O= O_A \otimes O_B$ is a observable in the world $W = (e^A_n \otimes e^B_k)$ of the composite system $\mathcal{Q}_A \times \mathcal{Q}_B.$

Generally, we can consider observable in a world of multipartite systems in a similar way.

\subsection{State}

To define the concept of a state in a world $W = (e_n),$ we denote by $\mathrm{vN} (W)$ the von Neumann subalgebra of $\mathcal{B} ( \mathbb{H})$ generated by all bounded operators in $\mathcal{O} (W),$ where $\mathcal{B} (\mathbb{H})$ is the von Neumann algebra of all bounded linear operators on $\mathbb{H}.$ A (pure) {\it state} in $W$ is then defined to be a (pure) state on $\mathrm{vN} (W).$ Recall that a state on a von Neumann algebra $\mathcal{M}$ is a linear functional $\varphi$ on $\mathcal{M}$ such that $\varphi (I) =1$ and $\varphi (T) \ge 0$ whenever $T \ge 0$ (whenever $T$ is a positive operator) (cf. \cite{Takesaki1979}). The set of states on $\mathcal{M}$ is a convex subset of the dual space of $\mathcal{M}$ which is compact in the $w^*$-topology. An extreme point of this convex set is called a pure state on $\mathcal{M},$ and we denote by $\mathcal{S} (\mathcal{M})$ the set of all pure states on $\mathcal{M}.$ By the Krein-Milman theorem (cf. \cite{Rudin1991}), this convex set is the closed convex hull of $\mathcal{S} ( \mathcal{M}).$

For a world $W = (e_n),$ the (vector) state $\omega_n,$ defined by $\omega_n (T) = \langle e_n | T | e_n \rangle,$ is a pure state in $\mathrm{vN} (W).$ Every $\omega_n$ defines a pure state of $W=(e_n)$ corresponding the pure state $| e_n \rangle$ in the conventional quantum mechanics. Each pure state $\omega_n$ can be shown to be completely determined by its values on $\mathrm{vN} (W),$ that is to say that the state $\omega_n$ has a unique extension to $\mathcal{B} (\mathbb{H})$ (cf. \cite{KS1959}). But there are many other pure states of $\mathrm{vN} (W).$

Indeed, the family of all pure states of $\mathrm{vN} (W)$ with the $w^*$-topology is the Stone-\u{C}ech compactification of $(\omega_n),$ which is homeomorphism to $\beta (\mathbb{N}),$ the Stone-\u{C}ech compactification of the integers. That is, $\mathcal{S} (\mathrm{vN} (W)) \cong \beta (\mathbb{N}),$ and so $\mathcal{S} (\mathrm{vN} (W)) \backslash \{ \omega_n \} \not= \emptyset.$ Those pure states other than $\omega_n$'s are mathematically obtained by a topological compactification, so we call these states {\it topology-compact states} in order to distinguish them from all usual states $\omega_n$'s which can be described by unit vectors in the associated Hilbert spaces. Such a topology-compact state must be determined by the totality of the associated world, that is all $e_n$'s. This is contrary to the fact that any vector state $\omega_k$ which is independent of the choice of the basis $(e_n)$ ($n \not= k$) for the ortogonal complement of $e_k,$ and is hence determined uniquely by $e_k$ alone.

Do those topology-compact states have unique extensions on $\mathcal{B} (\mathbb{H})$? This is a long-standing open problem posed by R. Kadison and I. Singer in their 1959 paper \cite{KS1959}, well known as the Kadison-Singer problem in mathematics. Recently, A. W. Marcus, D. A. Spielman and N. Srivastava \cite{MSS2015} have given a positive solution to the Kadison-Singer problem. This powerful result of Marcus, Spielmam and Srivastava implies that although a topology-compact state must be mathematically defined in a certain world, it also has absolute meaning as the same as a vector state.

We note that the topology-compact states do not appear in the usual formulation of quantum mechanics, of which any pure state is described by a unit vector in a Hilbert space. Since the topology-compact states are mathematically defined based the notion of a world, it is interesting to know whether or not could they be physically realized? Experimentally realization of such a topology-compact state will verify the necessity of the notion of a world introduced for a quantum system.

\subsection{Measurement}

Our description of a quantum system in terms of worlds is realistic, and so it describes certainties, not probabilities. There is indeterminateness only for observers if any. When making observation, that is measurement, we can apply the Copenhagen interpretation to a world for viewing a quantum state as an external observation, and obtain the Born rule of random outcomes.

In a certain world $W=(e_n)$ of a system $\mathcal{Q},$ a pure state $\omega$ belonging to $\mathcal{S} (W)$ gives maximal information that external observers can obtain by specifying the probabilities of the results of various observation which can be made on the system. In this case, every observable $O$ relative to this world takes a definite value $\omega ( O),$ and if $\omega = \omega_n,$ then $\omega (O)$ is the eigenvalue of $O$ corresponding to eigenvector $e_n,$ which is the same as the usual formulation of quantum mechanics.

However, if $O'$ is an observable relative to another world $W',$ different from $W,$ a measurement on it will give random outcomes, due to the fact that the information represented by that $\omega$ is obtained when the system is at the world $W$ and so the information is incomplete for $W'.$ By the powerful theorem of Marcus, Spielman, and Srivastava, $\omega$ has a unique extension to $\mathcal{B} (\mathbb{H}),$ still denote by $\omega$ this extension. Then, the expectation of measurement on $O'$ at $\omega$ is
\begin{equation}\label{eq:ExpValue}
\langle O' \rangle_\omega = \omega (O').
\end{equation}
In particular, if $\omega = \omega_n$ is a vector state described by $e_n,$ and if $O'$ is an observable relative to a world $W' = (e'_k)$ with a spectral decomposition $O' = \sum_k \lambda_k |e'_k \rangle \langle e'_k |,$ then the expectation of measurement on $O'$ is
\begin{equation}\label{eq:BornRule}
\omega_n (O') = \langle e_n | O' |e_n \rangle = \sum_k | \langle e_n | e'_k \rangle |^2 \lambda_k,
\end{equation}
and hence, the probability of $O'$ taking $\lambda_k$ is $| \langle e_n | e'_k \rangle |^2.$ This coincides with the usual Born rule of random outcomes on a vector state.

Now we return to the process of observation or making measurement. Given a state $\omega$ representing the information obtained by an observer when the system is at a world $W = (e_n),$ for observing an observable $O'$ relative to another world $W' = (e'_n),$ different from $W,$ we need to evolve the system from $W$ to $W'.$ Thanks to the fact that $O'$ is relative to $W',$ $O'$ takes a definite value $\omega' (O')$ in $W',$ which will be the outcome of making measurement. Note that $O'$ may be diagonalised under more than one orthonormal bases. The relativity of an observable requires that the observer first needs to determine the world that he or she involves for making measurement, and subsequently, the observer allows the unitary evolution of the system from $W$ to $W'$ given that the world $W'$ is chosen. Thus, the same $O'$ may take different values in different worlds, and the values are predetermined before observation.

Since every observable relative to $W'$ has a definite value, we can determine a quantum state $\omega'$ representing the information obtained by observer at $W'.$ By the theorem of Marcus, Spielman, and Srivastava again, $\omega'$ has a unique extension to $\mathcal{B} (\mathbb{H})$ yet. The expectation of subsequent measurement will depend only on the present $\omega',$ and is independent of the previous $\omega.$ That is to say, the change of information possessed by observer satisfies the Markovian property. Consequently, there exists the change of information in the process of observation, but no wavepacket collapse occurs as in the conventional formulation.

We note that for a topology-compact state $\omega$ of the world $W = (e_n),$ one has $\omega (|e_n \rangle \langle e_n|) = 0$ for any $n.$ Indeed, as pointed out in \cite{KS1959}, the topology-compact states are precisely those which annihilate all compact operators in $\mathcal{O} (W).$ On the other hand, it is well known that vector state are all  unitarily equivalent. However, as noted in \cite{KS1959}, this is not the case for topology-compact states. In fact, there are $2^{\mathfrak{\aleph}}$ inequivalent topology-compact states for $W,$ where $\aleph$ is the cardinality of $\mathbb{R}.$ Unfortunately, we have no explicit expression for topology-compact states. In Section \ref{topologystate} below, we will give a method of constructing topology-compact states and derive the corresponding Born rule.

\section{Non-locality}\label{nonlocality}

Nonlocality is a quantum mechanical property which was first described in 1935 by Einstein et al. \cite{EPR}. The EPR argument is best described in the Bell setup  of two electrons \cite{Bell1964}. Precisely, suppose that Alice and Bob in two separate sites have two spin $\frac{1}{2}$ particles with everyone having one particle, and the two-particle system is in the Bell state
\begin{equation}\label{eq:GAZstate}
| \Psi \rangle = \frac{1}{\sqrt{2}} \big ( | \uparrow_z \rangle_A | \uparrow_z \rangle_B + e^{\mathrm{i} \pi /4} | \downarrow_z \rangle_A | \downarrow_z \rangle_B \big ).
\end{equation}
Therefore, the measurement of the spin in each site and in any direction can be performed. Nonlocality arises if and only if we assume that the measurement of the spin of a particle collapses the Bell state from the linear superposition to either $| \uparrow_z \rangle_A | \uparrow_z \rangle_B$ or $| \downarrow_z \rangle_A | \downarrow_z \rangle_B,$ that is to say, measuring the spin of particle $A$ would instantaneously fix the spin of particle $B$ and vice versa, even if the two particles were allowed to separate to large distances. To check the nonlocality of quantum mechanics, Bell designed an inequality, well known as the Bell inequality \cite{Bell1964}. Instead of the original Bell inequality, we use the Clauser-Horne-Shimony-Holt (CHSH) inequality \cite{CHSH1969}. In this setup it is enough to consider the spin values in two directions, $x$ and $y.$ Assuming that there is no action at a distance in Nature implies that the spin value could not have been changed by distant measurements, therefore it existed before. This follows that
\begin{equation}\label{eq:CHSH}
v(\sigma^A_x) v (\sigma^B_x) + v ( \sigma^A_x) v ( \sigma^B_y ) + v ( \sigma^A_y ) v (\sigma^B_x ) - v ( \sigma^A_y) v (\sigma^B_y ) \leq 2,
\end{equation}
where $v (A)$ denotes the value taken by an observable $A.$ However,
\begin{equation}\label{eq:CHSHviolation}
\langle \Psi | \sigma^A_x \sigma^B_x + \sigma^A_x \sigma^B_y + \sigma^A_y \sigma^B_x - \sigma^A_y \sigma^B_y | \Psi \rangle = 2 \sqrt{2} >2,
\end{equation}
which shows that the assumption that there are definite predictions for all these results is inconsistent with the prediction of quantum mechanics. This is a proof of nonlocality based on the violation of Bell's inequalities, since the conclusion seems be that actions (measurements in $x$ or $y$ directions) of Alice change the outcome of Bob's measurement performed immediately after, vice versa.

However, physics has no mechanism for nonlocal actions, as suspected by Einstein et al. \cite{EPR}. To remove the action at a distance, the bset option as noted in \cite{Tipler2014, Vaidman2015} is to reject a tacit assumption, necessary for the Bell's proof, that there is only one world. That is to say, nonlocality disappears when the many-worlds interpretation is adopted. In fact, believing in the MWI, Alice knows that her prediction is not universally true. It is true only in her particular world. She knows that there are parallel worlds in which Bob's outcome is different. This also applies to Bob's measurement. Hence, there is no definite outcome exists prior to the measurement which is frequently assumed in Bell-type arguments and so the inequality \eqref{eq:CHSH} cannot hold in general.

Different from \cite{Tipler2014,Vaidman2015} based on Everett's ideas of many worlds in the ``relative state" formulation, we can use the relativity of observables based on the notion of worlds for removing the action at a distance. Indeed, $\sigma^A_x \sigma^B_x,\; \sigma^A_x \sigma^B_y,\; \sigma^A_y \sigma^B_x,$ and $\sigma^A_y \sigma^B_y$ are respectively relative to four different worlds:
\begin{equation}\label{eq:xx}
\{|\uparrow_x \rangle_A |\uparrow_x \rangle_B, |\uparrow_x \rangle_A |\downarrow_x \rangle_B, |\downarrow_x \rangle_A |\uparrow_x \rangle_B, |\downarrow_x \rangle_A |\downarrow_x \rangle_B \},
\end{equation}
\begin{equation}\label{eq:xy}
\{|\uparrow_x \rangle_A |\uparrow_y \rangle_B, |\uparrow_x \rangle_A |\downarrow_y \rangle_B, |\downarrow_x \rangle_A |\uparrow_y \rangle_B, |\downarrow_x \rangle_A |\downarrow_y \rangle_B \},
\end{equation}
\begin{equation}\label{eq:yx}
\{|\uparrow_y \rangle_A |\uparrow_x \rangle_B, |\uparrow_y \rangle_A |\downarrow_x \rangle_B, |\downarrow_y \rangle_A |\uparrow_x \rangle_B, |\downarrow_y \rangle_A |\downarrow_x \rangle_B \},
\end{equation}
and
\begin{equation}\label{eq:yy}
\{|\uparrow_y \rangle_A |\uparrow_y \rangle_B, |\uparrow_y \rangle_A |\downarrow_y \rangle_B, |\downarrow_y \rangle_A |\uparrow_y \rangle_B, |\downarrow_y \rangle_A |\downarrow_y \rangle_B \}.
\end{equation}
Hence, the inequality \eqref{eq:CHSH} should not hold in quantum regime because it is based on the assumption that all observables are in the same world. We see that removing the action at a distance by the relativity of observables is more explicit than that by the relativity of states.

Our argument also applies to the GHZ setup \cite{GHZ1989}. It only needs to translate the argument of Vaidman \cite{Vaidman2015} to our formalism. Thus, our reformulation excludes that quantum mechanics requires some ``spooky action at a distance", as shown in the conventional formulation of non-locality, such as the EPR paradox and various Bell's theorems. We should emphasize that our argument is based on the relativity of observables, while that of \cite{Tipler2014,Vaidman2015} on the relativity of states.

\section{Mathematical construction of topology-compact states}\label{topologystate}

Now, we present a mathematical method of constructing topology-compact states. This is done by using the so-called Banach limit. We denote by $\ell^{\infty} (\mathbb{N})$ the Banach space of all bounded sequence of complex numbers. A Banach limit is a continuous linear functional $L: \ell^{\infty} (\mathbb{N}) \mapsto \mathbb{C}$ such that for any sequences $x=(x_n)_{n \ge 1}$ and $y=(y_n)_{n \ge 1},$ the following conditions are satisfied:
\begin{enumerate}[{\rm i)}]

\item for any complex numbers $a, b,$ $L (a x + b y) = a L(x) + b L(y);$

\item if $x_n \ge 0$ for all $n,$ then $L(x) \ge 0;$

\item $L(x) = L(Sx),$ where $(Sx)_1=0$ and $(Sx)_{n+1} = x_n$ for all $n;$

\item if $x$ is a convergent sequence, then $L(x) = \lim_n x_n.$

\end{enumerate}
In other words, a Banach limit extends the usual limits. However, there exist sequences for which the values of two Banach limits do not agree.

Let $W = (e_n)$ be a world. Given a fixed Banach limit $L,$ for any operator $O$ in $\mathrm{vN} (W)$ with a spectral decomposition $O = \sum_n \lambda_n | e_n \rangle \langle e_n |,$ we define
\begin{equation}\label{eq:TopoStateBanach}
\omega_L (O) = L ( ( \lambda_n)_{n \ge 1} ).
\end{equation}
By definition, $\omega_L$ is a state on $\mathrm{vN} (W).$ Moreover, if $O$ is a compact operator, then $|\lambda_n| \to 0$ as $n \to \infty,$ and so $\omega_L (O) =0.$ Thus $\omega_L$ is a topology-compact state, which is pure whenever $L$ is an extreme point in the unit ball of $\ell^{\infty} (\mathbb{N})^*,$ the dual space of $\ell^{\infty} (\mathbb{N}).$

Now we derive the Born rule for $\omega_L.$ For an observable $O'$ relative to another world $W',$ by the Hahn-Banach theorem (cf. \cite{Rudin1987}), if we define
\begin{equation}\label{eq:TopoStateBorn}
\omega_L (O') = \inf_{(\lambda_n)_{n \ge 1} \in \ell^{\infty} (\mathbb{N}, \mathbb{R})} \Big ( L ((\lambda_n)_{n \ge 1}) + \Big \| \sum_{n \ge 1} \lambda_n | e_n \rangle \langle e_n | - O' \Big \| \Big )
\end{equation}
then $\omega_L$ can be extended to the space generated by $\mathrm{vN} (W)$ and $O'.$ On the other hand, by the theorem of Marcus, Spielman, and Srivastava, $\omega_L$ has a unique extension to $\mathcal{B} (\mathbb{H}).$ Hence, the expression on the right hand side of \eqref{eq:TopoStateBorn} is the uniquely possible value taken by $\omega_L$ on $O',$ which should be considered as the expectation of measurement on $O'$ at $\omega_L.$ Thus we obtain the Born rule for $\omega_L.$

Generally, for a given topology-compact state $\omega \in \mathcal{S} (W),$ by the theorem of Marcus, Spielman, and Srivastava we have
\begin{equation}\begin{split}
\inf_{(\lambda_n)_{n \ge 1} \in \ell^{\infty} (\mathbb{N}, \mathbb{R})} & \Big ( \omega \Big ( \sum_n \lambda_n | e_n \rangle \langle e_n | \Big ) + \Big \| \sum_{n \ge 1} \lambda_n | e_n \rangle \langle e_n | - O' \Big \| \Big )\\
= & \sup_{(\lambda_n)_{n \ge 1} \in \ell^{\infty} (\mathbb{N}, \mathbb{R})} \Big ( \omega \Big ( \sum_n \lambda_n | e_n \rangle \langle e_n | \Big ) - \Big \| \sum_{n \ge 1} \lambda_n | e_n \rangle \langle e_n | - O' \Big \| \Big )
\end{split}\end{equation}
for any observable $O'$ relative to another world $W',$ and hence can take
\begin{equation}
\omega (O') = \sup_{(\lambda_n)_{n \ge 1} \in \ell^{\infty} (\mathbb{N}, \mathbb{R})} \Big ( \omega \Big ( \sum_n \lambda_n | e_n \rangle \langle e_n | \Big ) - \Big \| \sum_{n \ge 1} \lambda_n | e_n \rangle \langle e_n | - O' \Big \| \Big )
\end{equation}
as the expectation of $O'$ at $\omega.$ This is the Born rule for any topology-compact state $\omega \in \mathcal{S} (W).$

\section{Conclusion}\label{Conclusion}

In conclusion, we have formulated a mathematical formalism of quantum mechanics based on the notion of a world, in which the concepts of observable and state are both derived notions. A quantum system is completely determined by a world of it. Yet, the evolution of the system is described by the Schr\"{o}diner equation for the worlds of it. Any observable must be considered in a certain world and hence is a notion of relativity. However, a state defined in a certain world has a unique extension to the total system and so is of absolute meaning. In particular, the so-called topology-compact states appear naturally in a world except for vector states in the usual formulation of quantum mechanics. By applying the Copenhagen interpretation to a world for regarding a quantum state as an external observation, indeterminateness of outcomes appears for the observer making observation, and the Born rule of random outcomes can be obtained. We give a mathematical method of constructing topology-compact states and derive the Born rule for them. In our formalism of quantum mechanics, the action at a distance disappears naturally from the relativity of observables.

It is emphasized that contrary to the usual formulation of the many-worlds interpretation \cite{Vaidman2014}, our notion of a world is a rigorously
defined  mathematical entity and so a realistic concept in the sense of \cite{HS2010}. This leads to the concept of a topology-compact state beyond vector states in the orthodox formalism of quantum mechanics. It is argued that experimentally realization of such a topology-compact state will verify the necessity of the notion of a world introduced for a quantum system.

Therefore, a mathematically complete theory of quantum mechanics should consist of two parts. The first part is the realistic description of a quantum system by the notion of a world, which is taken as the basic physical entity with no {\it a priori} interpretation. The evolution of the system is governed by the Schr\"{o}dinger equation for the worlds of it, and hence there is no indeterminism in this part. The second part is the statistical description for the quantum system under consideration by the notion of a quantum state, that is the Copenhagen interpretation for the standard quantum mechanics in which probability appears due to incompleteness of information. In this part, there exists only the change of information satisfying the Markovian property, and thus there are various uncertainty relations between quantum variables as formulated in the conventional formalism of quantum mechanics.

\

This work was partially supported by the Natural Science Foundation of China under Grant No.11171338 and No.11431011, and National Basic Research Program of China under Grant No. 2012CB922102.

\bibliography{apssamp}

\end{document}